\documentclass{midl} 

\usepackage{graphicx,wrapfig,lipsum}
\jmlrworkshop{MIDL 2019 -- Full paper track}
\title[Joint lesion and tissue segmentation from task-specific hetero-modal datasets]{Learning joint lesion and tissue segmentation from task-specific hetero-modal datasets}

\midlauthor{\Name{Reuben Dorent \nametag{$^{1}$}} \Email{reuben.dorent@kcl.ac.uk}\\
\Name{Wenqi Li \midlotherjointauthor\nametag{$^{1}$}} \Email{wenqi.li@kcl.ac.uk}\\
\Name{Jinendra Ekanayake \midlotherjointauthor\nametag{$^{2}$}}  \Email{j.ekanayake@ucl.ac.uk}\\
\Name{Sebastien Ourselin \midlotherjointauthor\nametag{$^{1}$}}  \Email{sebastien.ourselin@kcl.ac.uk}\\
\Name{Tom Vercauteren \midlotherjointauthor\nametag{$^{1}$}} \Email{tom.vercauteren@kcl.ac.uk}\\
\addr $^{1}$ School of Biomedical Engineering \& Imaging Sciences, King’s College London, London, UK \\
\addr $^{2}$ Wellcome / EPSRC Centre for Interventional and Surgical Sciences, UCL, London, UK
}

\begin{document}
\maketitle

\begin{abstract}
Brain tissue segmentation from multimodal MRI is a key building block of many neuroscience analysis pipelines. It could also play an important role in many clinical imaging scenarios.
Established tissue segmentation approaches have however not been developed to cope with large anatomical changes resulting from pathology. The effect of the presence of brain lesions, for example, on their performance is thus currently uncontrolled and practically unpredictable.
Contrastingly, with the advent of deep neural networks (DNNs), segmentation of brain lesions has matured significantly and is achieving performance levels making it of interest for clinical use.
However, few existing approaches allow for jointly segmenting normal tissue and brain lesions.
Developing a DNN for such joint task is currently hampered by the fact that annotated datasets typically address only one specific task and rely on a task-specific hetero-modal imaging protocol.
In this work, we propose a novel approach to build a joint tissue and lesion segmentation model from task-specific hetero-modal and partially annotated datasets.
Starting from a variational formulation of the joint problem, we show how the expected risk can be decomposed and optimised empirically.
We exploit an upper-bound of the risk to deal with missing imaging modalities.
For each task, our approach reaches comparable performance than task-specific and fully-supervised models.
\end{abstract}


\section{Introduction}
Traditional approaches for tissue segmentation used in brain segmentation software packages such as FSL \citep{Jenkinson:NeuroImage:2012}, SPM \citep{Ashburner:NeuroImage:2000} or NiftySeg \citep{Cardoso:TMI:2015} are based on subject-specific optimisation. FSL and SPM fit a Gaussian Mixture Model to the MR intensities using either a Markov Random Field (MRF) or tissue prior probability maps as regularisation. Alternatively, multi-atlas methods rely on label propagation and fusion from multiple fully-annotated images, i.e. atlases, to the target image \citep{Iglesias:MedIA:2015}. These methods typically require extensive pre-processing, e.g. skull stripping, correction of bias field or registration. They are also often time-consuming, and are inherently only adapted for brains devoid of large anatomical changes induced by pathology. Indeed, it has been showed that the presence of lesions distorts the registration output \citep{Sdika:HBM:2009}. Similarly, lesions introduce a bias in the MRF. This leads to a performance degradation in presence of lesions for brain volumes measurement \cite{Battaglini:HBM:2012} and any subsequent analysis.

While quantitative analysis is expected to play a key role in improving the diagnosis and follow-up evaluations of patients with brain lesions, current tools mostly focus on the lesions themselves.
Existing quantitative neuroimaging approaches allow the extraction of imaging biomarkers such as the largest diameter, volume, and count of the lesions. Thus, automatic segmentation of the lesions promises to speed up and improve the clinical decision-making process but more refined analysis would be feasible from tissue classification and region parcellation.
As such, although very few works have addressed this problem yet, a joint model for lesion and tissue segmentation is expected to bring significant clinical impact.

Deep Neural Networks (DNNs) became the state-of-the-art for most of the segmentation tasks and one would now expect to train a joint lesion and tissue segmentation algorithm. Yet, DNNs require a large amount of annotated data to be successful. Existing annotated databases are usually task-specific, i.e. providing either scans with brain tissue annotations for patients/controls devoid of large pathology-induced anatomical changes, or lesion scans with only lesion annotations. For this reason, the imaging protocol used for the acquisition also typically differs from one dataset to another. Indeed, tissue segmentation is usually performed on T1 scans, unlike lesion segmentation which normally also encompasses Flair \cite{Barkhof:CerebDis:2002}. Similarly, the resolution and contrast among databases may also vary.
Given the large amount of resources, time and expertise required to annotate medical images, given the varying imaging requirement to support each individual task and given the availability of task-specific databases, it is unlikely that large databases for every joint problem, such as lesion and tissue segmentation, will become available for research purposes. There is thus a need to exploit task-specific databases to address joint problems. Learning a joint model from task-specific hetero-modal datasets is nonetheless challenging. This problems lies at the intersection of Multi-Task Learning, Domain Adaptation and Weakly Supervised Learning with idiosyncrasies making individual methods from these underpinning fields insufficient to address it completely.

Multi-Task Learning (MTL)  aims  to  perform  several tasks simultaneously by extracting some form of common knowledge or representation and introducing a task-specific back-end. When relying on DNN for MTL, usually the first layers of the network are shared, while the top layers are task-specific. The global loss function is often a sum of task-specific loss functions with manually tuned weights. Recently, \citet{Kendall:NIPS:2017} proposed a Bayesian parameter-free method to estimate the MTL loss weights and \citet{Bragman:MICCAI:2018} extended it to spatially adaptive task weighting and applied it to medical imaging.
In addition to arguably subtle differences between MTL and joint learning discussed in more depth later, MTL approaches do not provide any mechanism for dealing with hetero-modal datasets and changes in imaging characteristics across task-specific databases.

Domain Adaptation (DA) is a solution for dealing with heterogeneous datasets. The main idea is to create a common feature space for the two sets of scans. \citet{Csurka:DACVARevChap:2017} proposed an extensive comparison of these methods in deep learning. Learning from hetero-modal datasets could be consider as a particular case of DA. \citet{Havaei:MICCAI:2016} proposed a network architecture for dealing with missing modalities. However, DA methods focus on solving a single task and rely on either fully-supervised approaches or unsupervised adaptation as done by \citet{Kamnitsas:MICCAI:2017}.

Weakly-supervised Learning (WSL) deals with missing, inaccurate, or inexact annotations. Our problem is a particular case of learning with missing labels since each dataset provide a set of labels and the two sets are disjoint. \citet{Li:PAMI:2017} proposed a method to learn a new task from a model trained on another task. This method combines DA through transfer learning and MTL. At the end, two models are created: one for the first task and one for the second one.  \citet{Kim:WACV:2018} extent this approach by using a knowledge distillation loss in order to create a unique joint model. This aims to alternatively learn one task without forgetting the other one. The WSL problem was thus decomposed into a MTL problem with similar limitations for our specific use case.

The contributions of this work are four-fold. First we propose a joint model that performs tissue and lesion segmentation as a unique joint task and thus exploits the interdependence between lesion and tissue segmentation tasks. Starting from a variational formulation of the joint problem, we exploit the disjointness of the label sets to propose a practical decomposition of the joint loss. Secondly, we introduce feature channel averaging across modalities to adapt existing networks for our hetero-modal problem. Thirdly, we develop a new method to minimise the expected risk under the constraint of missing modalities. Relying on reasonable assumptions, we show that the expected risk can be further decomposed and minimised via a tractable upper bound. To our knowledge, no such optimisation method for missing modalities in deep learning has been published before. Finally, we evaluate our framework for white matter lesions and tissue segmentation. We demonstrate that our joint approach can achieve, for each individual task, similar performance compared to a task-specific baseline.
Albeit relying on different annotation protocols, results using a small fully-annotated joint dataset demonstrate efficient generalisability.

\section{Tissue and lesion segmentation as a single task}
In order to develop a joint model, we propose a mathematical variational formulation of the problem and a method to optimise it empirically.

\subsection{Formal problem statement}
Let $\textbf{x}=(x^1,..,x^M) \in \mathcal{X}=\mathbb{R}^{N\times M}$ be a vectorized multimodal image and $y \in \mathcal{Y}=\{0,..,C\}^{N}$ its associated segmentation map. $N$, $M$ and $C$ are respectively the number of voxels, modalities and classes. Our goal is to determine a predictive function $h_{\theta}:\mathcal{X} \mapsto \mathcal{Y}$ that minimises the discrepancy between the ground truth label vector $y$ and the prediction $h_{\theta}(\textbf{x})$. Let $\mathcal{L}$ be a loss function that computes this discrepancy.
Following the formalism used by \citet{Bottou:SIAMRev:2018}, given a probability distribution $\mathcal{D}$ over $(\mathcal{X},\mathcal{Y})$ and random variables $(X,Y)$ under this distribution, we want to find $\theta ^{*}$ such that:
\begin{equation} \label{eq:0}
 \theta^{*}=\text{argmin}_{\theta} \mathbb{E}_{(X,Y) \sim \mathcal{D}}\left[\mathcal{L}\left(h_{\theta}(X), Y\right)\right] 
\end{equation}
Let $\mathcal{C}_t$, $\mathcal{C}_l$  and $0$ be respectively the tissue classes, the lesion classes and the background class. Since $\mathcal{C}_t$ and $\mathcal{C}_l$ are disjoint, the segmentation map $y$ can be decomposed into two segmentation maps $y_i = y^{l}_i+y^{t}_i$ with $y^{t}_i \in \mathcal{C}_t\cup \{0\}$, $y^{l}_i \in \mathcal{C}_l\cup \{0\}$, as shown in \figureref{fig:decomposition}.
\begin{figure}[tb!]
\floatconts
  {fig:decomposition}
  {\caption{Decomposition of the label map into the sum of two segmentation maps.}}
  {\includegraphics[width=\linewidth]{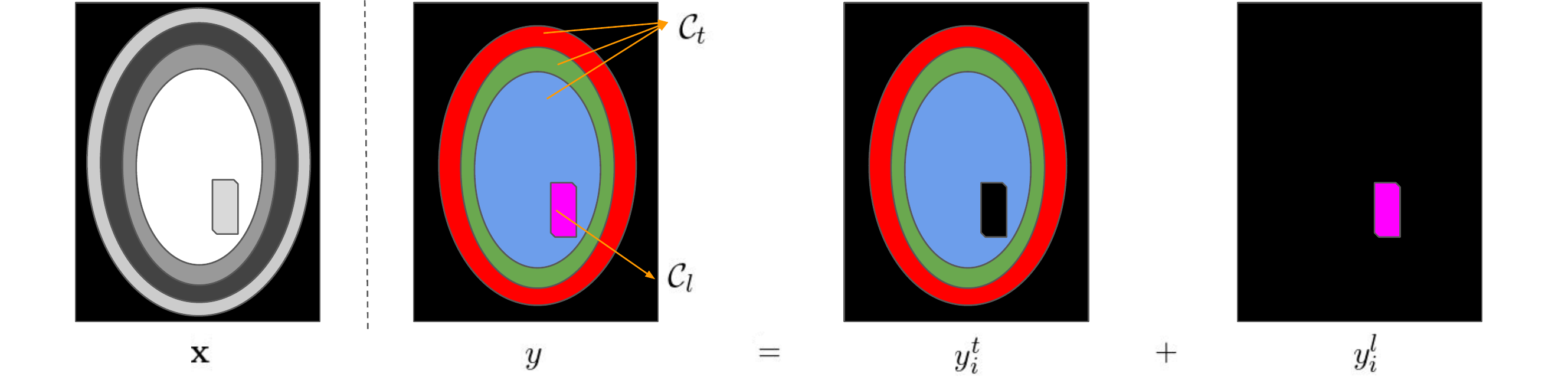}}
\end{figure}

Let's assume that the loss function $\mathcal{L}$ can also be decomposed into a tissue loss function $\mathcal{L}^{t}$ and a lesion loss function $\mathcal{L}^{l}$. This is common for multi-class segmentation loss functions in particular for those with \emph{one-versus-all} strategies (e.g. Dice loss, Jaccard loss):
\begin{equation}\label{eq:H1}\tag{$\textbf{H}_1$}
\mathcal{L}(h_{\theta}(X), Y) = \mathcal{L}^{t}(h_{\theta}(X), Y^{t}) + \mathcal{L}^{l}(h_{\theta}(X), Y^{l})
\end{equation}
Then, \equationref{eq:0} can be rewritten as:
\begin{equation} \label{eq:1}
\theta^{*} = \text{argmin}_{\theta}  \underbrace{\mathbb{E}_{(X,Y) \sim \mathcal{D}}[\mathcal{L}^{t}(h_{\theta}(X), Y^{t})]}_{\mathcal{R}^t(\theta)} + \underbrace{\mathbb{E}_{(X,Y) \sim \mathcal{D}}[\mathcal{L}^{l}(h_{\theta}(X), Y^{l})]}_{\mathcal{R}^l(\theta)}
\end{equation}

\subsection{On the distribution \texorpdfstring{$\mathcal{D}$}{D} in the context of heterogeneous databases}
As we expect different distributions across heterogeneous databases, two probability distributions of $(X,Y)$ over $(\mathcal{X},\mathcal{Y})$ can be distinguished: 1/ under $\mathcal{D}_{control}$, $(X,Y)$ corresponds to a multimodal scan and segmentation map of a patient without lesions. Note that although we use the term \emph{control} for convenience, we expect to observe pathology with "diffuse" anatomical impact, e.g. from dementia. 2/ under $\mathcal{D}_{lesion}$, $(X,Y)$  corresponds to a multimodal scan and segmentation map of a patient with lesions.

Since traditional methods are not adapted in the presence of lesions, the most important and challenging distribution $\mathcal{D}$ to address is the one for patients with lesions, $\mathcal{D}_{lesion}$.
In the remainder of this work we thus assume that:
\begin{equation}\label{eq:H2}\tag{$\textbf{H}_2$}
\mathcal{D} \triangleq \mathcal{D}_{lesion}.
\end{equation}

\subsection{Hetero-modal network architecture}
In order to learn from hetero-modal datasets, we need a network architecture that allows for missing modalities. We proposed an architecture inspired by HeMIS \citep{Havaei:MICCAI:2016} and HighResNet \citep{Li:IPMI:2017} shown in \figureref{fig:network}. Features of each modality are first extracted separately and are then averaged. The spatial resolution of the input and the output are the same. Dilated convolutions and residual connections are used to capture information at multiple scales and avoid the problem of vanishing gradients. This network with weights $\theta$ is used to capture the predictive function $h_{\theta}$.
\begin{figure}[tb!]
\floatconts
  {fig:network}
  {\caption{The proposed network architecture: a mix between HighResNet and HeMIS. To avoid cluttering, only one of the three convolution blocks is shown in the residual blocks.}}
  {\includegraphics[width=\linewidth]{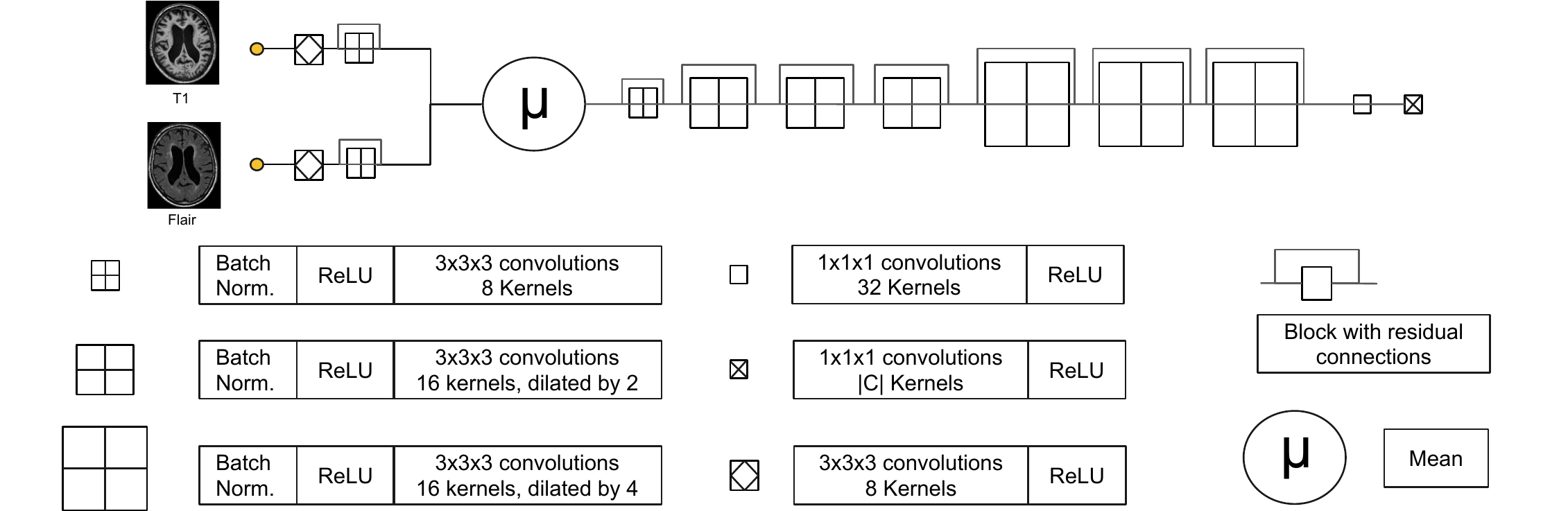}}
\end{figure}

\subsection{Upper-bound for the tissue expected risk \texorpdfstring{$\mathcal{R}^t$}{}}
Although thanks to its hetero-modal architecture, $h_{\theta}$ may now handle inputs with varying number of modalities, the current decomposition of \eqref{eq:0} assumes that all the modalities of $X$ are available for evaluating the loss. In our scenario, we have only access to T1 control scans with tissue annotations or T1 and Flair scans with only lesion annotations. Consequently, as we do not have any T1 and Flair images with tissue annotations, and as evaluating a loss with missing modalities would lead to a bias, estimating $\mathcal{R}^{t}$ is not straightforward. 

In this section we propose an upper-bound of $\mathcal{R}^{t}$ using T1 control images with tissue annotations and outputs from the network.
Let's assume that the loss function $\mathcal{L}^{t}$ satisfies the triangle inequality (e.g. Jaccard loss):
\begin{equation}\label{eq:H3}\tag{$\textbf{H}_3$}
\forall (a,b,c) \in \mathcal{Y}^{3}: \ \mathcal{L}^{t}(a,c) \leq \mathcal{L}^{t}(a,b) + \mathcal{L}^{t}(b,c)
\end{equation}
Let $p$ denote the projection of $\textbf{x}$ (will all the modalities) to the T1 modality, $p:\textbf{x}=(x^{\textrm{T1}},x^{\textrm{Flair}})\mapsto x^{\textrm{T1}}$. Under \eqref{eq:H3}, $\mathcal{L}^t$ satisfies the following inequality:
\begin{equation}
\mathcal{L}^{t}(h_{\theta}(X), Y^{t}) \leq \mathcal{L}^{t}(h_{\theta}(X), h_{\theta}(p(X))) + \mathcal{L}^{t}(h_{\theta}(p(X)), Y^{t})
\end{equation}
In combination with \eqref{eq:H2}, this leads to:
\begin{equation}\label{eq:upbnd_init}
    \mathcal{R}^{t}(\theta) \leq  \mathbb{E}_{(X,Y) \sim \mathcal{D}_{lesion}}[\mathcal{L}^{t}(h_{\theta}(X), h_{\theta}(p(X)))] + \mathbb{E}_{(X,Y) \sim \mathcal{D}_{lesion}}[\mathcal{L}^{t}(h_{\theta}(p(X)), Y^t)]
\end{equation}

The decomposition in \eqref{eq:upbnd_init} requires comparison of inference done from T1 inputs, i.e. $h_{\theta}(p(X))$ with ground truth tissue maps $Y^t$. While this provides a step towards a practical evaluation of $\mathcal{R}^{t}$, we still face the challenge of not having tissue annotations $Y^t$ under $\mathcal{D}_{lesion}$.
Let us further assume that the restriction of the distributions $\mathcal{D}_{lesion}$ and $\mathcal{D}_{control}$ to the parts of the brain not affected by lesions are the same, i.e.:
\begin{equation}\label{eq:H4}\tag{$\textbf{H}_4$}
\forall i \in\{1...N\} P_{ \mathcal{D}_{lesion}}(x_i,y_i|y_i \in \mathcal{C}_{tissue})=P_{ \mathcal{D}_{control}}(x_i,y_i|y_i \in \mathcal{C}_{tissue})
\end{equation}
By combining \eqref{eq:H3} and \eqref{eq:H4}, an upper bound of $\mathcal{R}^{t}$ can be provided as:
\begin{equation}
    \label{eq:upper}
    \mathcal{R}^{t}(\theta) \leq  \underbrace{\mathbb{E}_{(X,Y) \sim \mathcal{D}_{lesion}}[\mathcal{L}^{t}(h_{\theta}(X), h_{\theta}(p(X)))]}_{\mathcal{R}^{t}_{1}(\theta)} + \underbrace{\mathbb{E}_{(X,Y) \sim \mathcal{D}_{control}}[\mathcal{L}^{t}(h_{\theta}(p(X)), Y^t)]}_{\mathcal{R}^{t}_{2}(\theta)}
\end{equation}
As observed in \eqref{eq:upper}, the upper-bound is the sum of the expected loss between the T1 scan outputs and the labels and the expected loss between the outputs using either one or two modalities as input. We emphasise that, to the best of our knowledge, this second loss term does not appear in existing heteromodal approaches such as HeMIS \citep{Havaei:MICCAI:2016}.

\subsection{Empirical estimation of the decomposed loss}

\begin{wrapfigure}{r}{0.50\textwidth}
  \centering
  \includegraphics[width=0.46\textwidth]{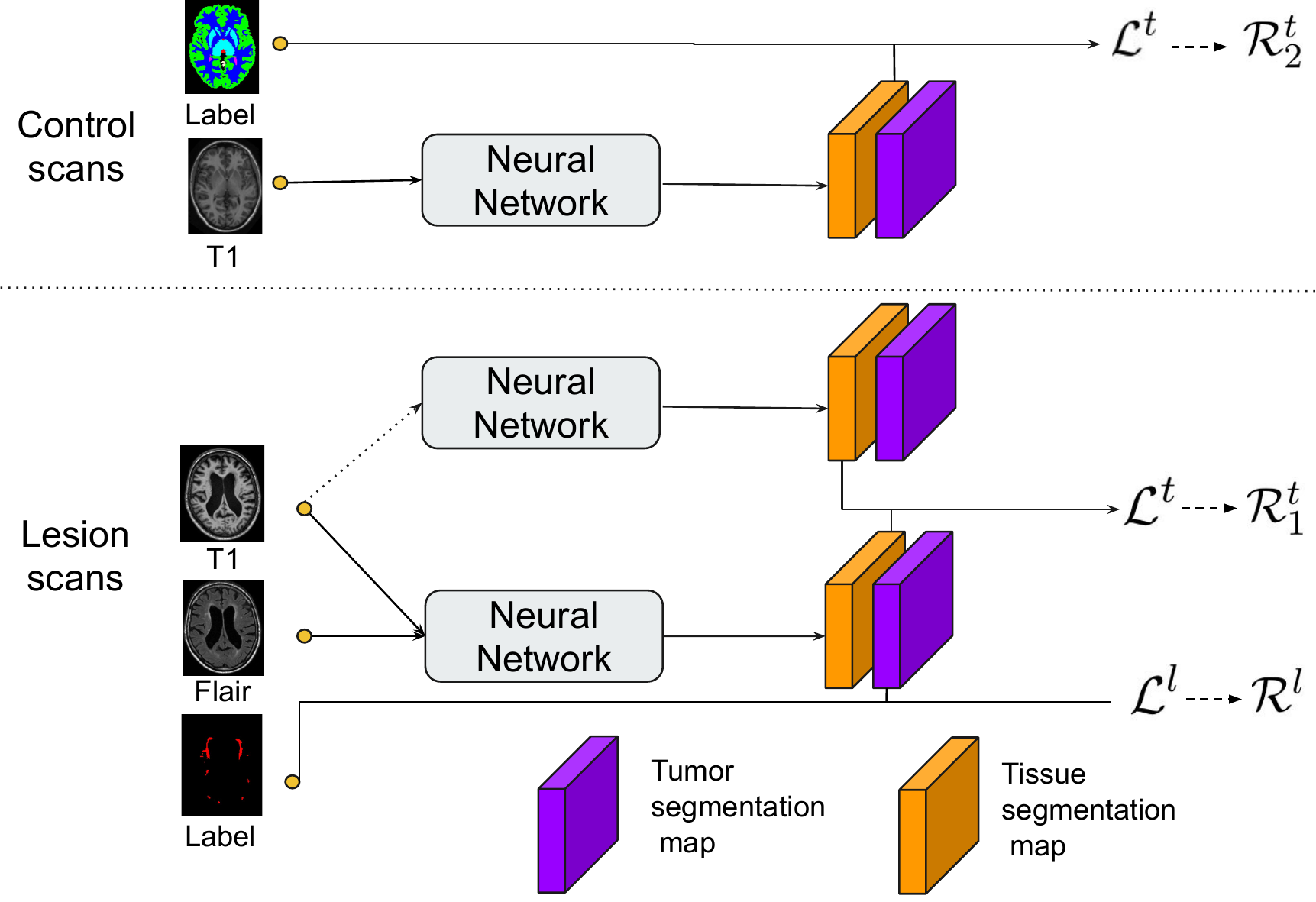}
  \caption{Procedure for estimating the expected risks $\mathcal{R}^l$, $\mathcal{R}^{t}_{1}$ and $\mathcal{R}^{t}_{2}$.}
  \label{fig:optimisation}
\end{wrapfigure}

As is the norm in data-driven learning, we do not have access to the true joint probabilities 
$\mathcal{D}_{control}$ or $\mathcal{D}_{lesion}$.
The common method is to estimate the expected risk using training samples. In our case, we have two hetero-modal training samples $\mathcal{S}_{control}$ and $\mathcal{S}_{lesion}$ with respectively tissue and lesion annotations. We can estimate the expected risks $\mathcal{R}^l(\theta)$, $\mathcal{R}^{t}_{1}(\theta)$, $\mathcal{R}^{t}_{2}(\theta)$ by respectively using lesion segmentation outputs of lesion T1+Flair scans, tissue segmentation outputs from T1 and T1+Flair scans and tissue segmentation outputs of control T1 scans. \figureref{fig:optimisation} illustrates the complete training procedure.

\section{Experiments}
While focusing on the white matter lesion and tissue segmentation problem, our goal in the following experiments is to predict six tissue classes (white matter, gray matter, basal ganglia, ventricles, cerebellum, brainstem), the white matter lesions and the background.

\subsection{Data}
To demonstrate the feasibility of our joint learning approach, we used three sets of data.

\paragraph{Lesion data $S_{lesion}$:} The White Matter Hyperintensities (WMH) database consists of 60 sets of brain MR images (T1 and Flair, $M=2$) with manual annotations of WMH (\url{http://wmh.isi.uu.nl/}). The data comes from three different institutes.

\paragraph{Tissue data $S_{control}$:} Neuromorphometrics provided 32 T1 scans ($M^{'}=1)$ for MICCAI 2012 with manual annotations of 155 structures of the brain from which we deduct the six tissue classes. In order to have balance training datasets for the two types of segmentation, and similar to \citet{Li:IPMI:2017}, we added 28 T1 control scans from the ADNI2 dataset with bronze standard parcellation of the brain structures computed with the accurate but time-consuming algorithm of \citet{Cardoso:TMI:2015}.

\paragraph{Fully annotated data:} MRBrainS18 (\url{http://mrbrains18.isi.uu.nl/}) is composed of 30 sets of brain MR images with tissue and lesions manual annotations. Only 7 MR images are publicly available. We used this data only for evaluation and not for training. To be consistent with the lesion data, the cerebrospinal fluid is considered as background. 

To satisfy the assumption \eqref{eq:H4}, we resampled the data to $1 \times 1 \times3$ mm\textsuperscript{3}, used a histogram-based scale \cite{Milletari:3DV:2016} and a zero-mean unit-variance normalization.

\subsection{Choice of the loss}
We used the probabilistic version of the Jaccard loss for $\mathcal{L}$:
\begin{equation}
\mathcal{L}(h_{\theta}(\textbf{x}),y) = 1-\sum_{c \in C} \omega_{c} \frac{\sum_{j=1}^N g_{j,c}p_{j,c}}{\sum_{j=1}^N g_{j,c}^2 + p_{j,c}^2 - g_{j,c}p_{j,c} } \ \ \ \text{such as } \sum_{c \in C} \omega_{c} =1
\end{equation}
\eqref{eq:H1} is satisfied because of the \emph{one-versus-all} strategy, i.e. sum over the classes of a class-specific loss. In order to give the same weight to the lesion segmentation and the tissue segmentation, we choose for any tissue class $c$, $w_c= \frac{1}{16}$ and for the lesion class $l$, $w_l= \frac{1}{2}$. While the triangle inequality holds for the Jaccard distance \cite{Kosub:PRL:2018}, the proof that its probabilistic version also satisfies it, i.e. \eqref{eq:H3}, is left for future work.

\subsection{Implementation details}
We implemented our network in NiftyNet, a Tensorflow-based open-source platform for deep learning in medical imaging \cite{Gibson:CMPB:2018}. Convolutional layers are initialised such as \citet{He:ICCV:2015}. The scaling and shifting parameters in the batch normalisation layers were initialised to 1 and 0 respectively. As suggested by \cite{Ulyanov:arXiv:2016}, we used instance normalization for inference. We used the Adam optimisation method \citep{Kingma:arXiv:2014}. The learning rate $l_R$, $\beta_1$, $\beta_2$ were respectively set up to 0.005, 0.9 and 0.999. At each training iteration, we feed the network with one image from the tissue dataset and one from the lesion dataset. $120\times120\times40$ sub-volumes were randomly sampled from the training data using an uniform sampling for the tissue data and a weighted sampling based on dilated lesions maps for the lesion data. The models were trained until we observed a plateau in performance on the validation set. We experimentally found that the inter-modality loss has to be skipped for the first (5000) iterations. We randomly spitted the data into $70\%$ for training, $10\%$ for validation and $20\%$ for testing for each of the 4 folds.
\begin{figure}[tb!]
\floatconts
  {fig:results}
  {\caption{Segmentation results using our method and task-specific models. (1) axial slice from test image volumes from (a) WMH and (b) Neuromorphometrics, (2) manual annotations, (3) outputs from the joint learning model, (4) outputs from the tissue segmentation (\textit{N}) model, (5)  outputs from the lesion segmentation (\textit{W}) model }}
  {\includegraphics[width=0.7\linewidth]{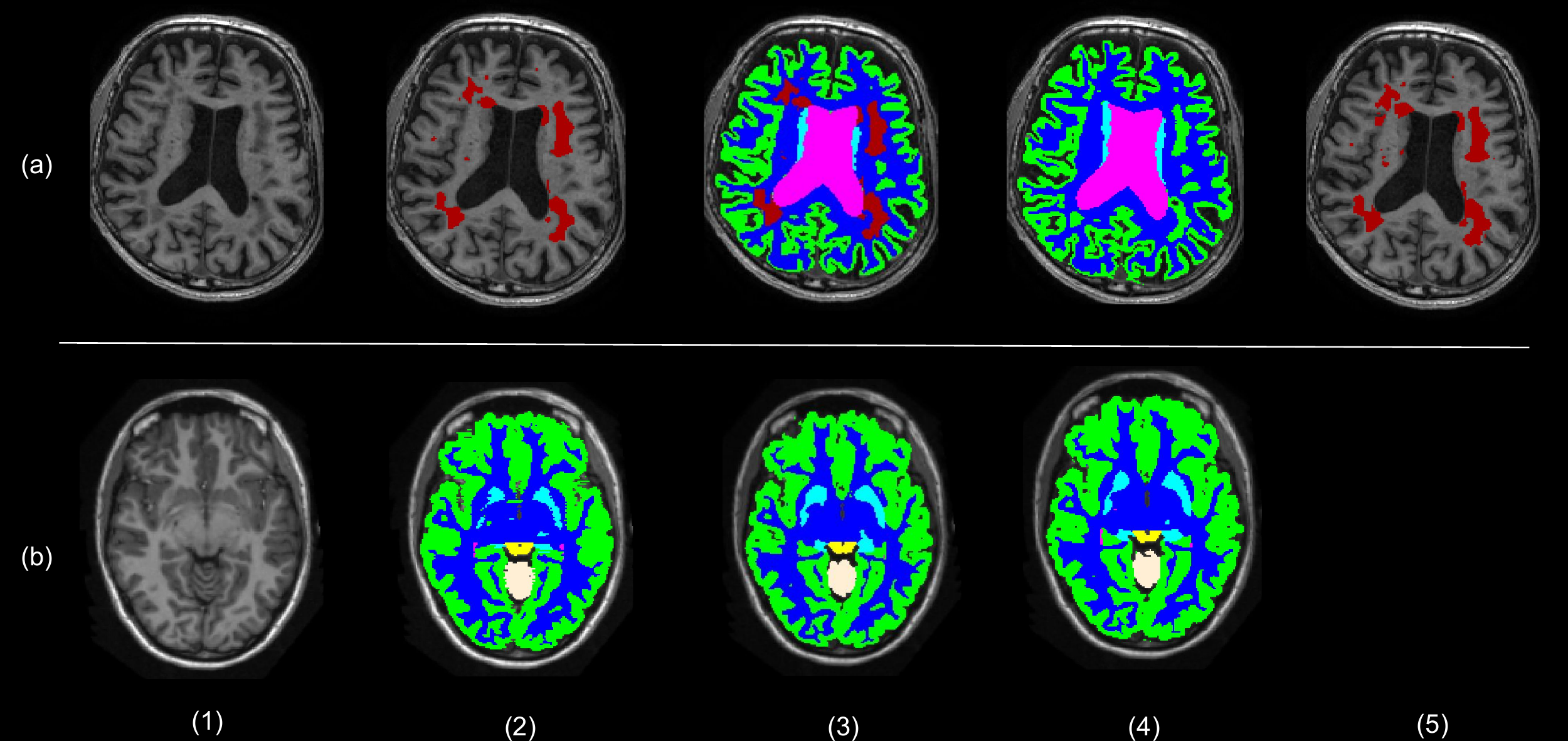}}
\end{figure}

\subsection{Results for the joint learning model}
\paragraph{Joint learning versus single task learning}
First, we compare individual models to the joint model using our approach. The lesion segmentation (\textit{W}) model was trained on WMH dataset with the lesion annotations, the tissue segmentation (\textit{N}) on Neuromorphometrics dataset with the tissue annotations, and our method (\textit{W+N}) on WMH and Neuromorphometric datasets with their respective set of annotations. The similarity between the prediction and the ground truth is computed using the Dice Similarity Coefficient (DSC) for each class. \tableref{tab:scores} and \figureref{fig:results} show the results of these models on test images. The joint model and single task models achieve comparable performance. This suggest that learning from hetero-modal datasets via our method does not degrade the task-specific performance. Moreover, we observe in \figureref{fig:results} that the tissue knowledge learnt from T1 scans has been well generalised to multi-modal scans.

\begin{table}[h]
\floatconts
  {tab:scores}%
  {\caption{Comparison between the lesion segmentation model \textit{W}, the tissue segmentation model \textit{N}, the fully-supervised model (\textit{M}), a traditional approach  (\textit{SPM}) and our joint model (\textit{W+N}). Dice Similarity Coefficients ($\%$) has been computed.}}%
  {\begin{tabular}{c|ccc|ccc|cccc}
 &  \multicolumn{3}{c|}{Neuromorphometrics} & \multicolumn{3}{c|}{WMH} & \multicolumn{3}{c}{MRBrainS18}  \\
    & \textit{N} & \textit{M} & \textit{W+N} & \textit{W} &  \textit{M} & \textit{W+N}  & \textit{SPM} & \textit{M} & \textit{W+N}   \\
    \hline 
Gray matter &  88.5 & 42.0 & 89.4 & & & & 76.5 & 83.3 & 79.4\\
White matter & 92.4 & 56.7 & 92.8 & & & & 75.7 & 85.9 & 85.4 \\
Brainstem & 93.4 & 20.0 & 93.1& & & & 76.5 & 92.3 & 72.3 \\
Basal ganglia & 86.7 & 41.2 & 87.2  & & & & 74.7 & 79.1 & 75.3   \\
Ventricles &  90.7 & 24.5 & 91.6  & & & &  80.9 & 91.0 & 91.7 \\
Cerebellum &  92.5 & 43.7 & 94.9 & & & & 89.4 & 91.8 & 90.8  \\
White matter lesion &  & & & 61.9 & 50.6 & 59.9  & 40.8 &53.5 & 53.7  \\
\end{tabular}}
\end{table}

\paragraph{Joint model versus fully-supervised model}
In this section, we compare our method (\textit{W+N}) to the fully-supervised (\textit{M}) model trained on MRBrainS18 using both tissue and lesion annotations. We evaluated the performance on the three datasets. On the one hand, we submitted our models to the online challenge MRBrainS18. One of the major benefits of evaluating our method on a challenge is to directly benchmark our method with existing methods, in particular with traditional methods such as SPM \citep{Ashburner:NeuroImage:2000}. On the other hand, we compared the performance on the tissue and lesion datasets using either all the scans (\textit{M}) or the testing split (\textit{W+N}). The DSC was computed for each class and \tableref{tab:scores} show the results. First our model outperforms SPM on 6 of the 7 classes. Secondly, the two models achieve very similar performance on lesion segmentation. Concerning the tissue segmentation, as expected, each of the network outperforms on its training datasets. However, the fully supervised model doesn't generalise to Neuromorphometrics dataset. In contrast, the differences are smaller for the tissue segmentation classes on MRBrainS18. Especially, \figureref{fig:differences} shows differences in the annotation protocol between MRBrainS18 and Neuromorphometrics for the white matter, brainstem and cerebellum and how it affects the predictions.

\begin{figure}[tb!]
\floatconts
  {fig:differences}
  {\caption{Annotation protocol comparison between scans from (a) Neuromorphometrics and (b) MRBrainS18. (1) sagital slice from test images volumes, (2) manual annotations, (3) outputs from our method \textit{(W+N)}, (4) outputs from fully-supervised model \textit{(N)}. Arrows show the protocol differences.}}
  {\includegraphics[width=0.7\linewidth]{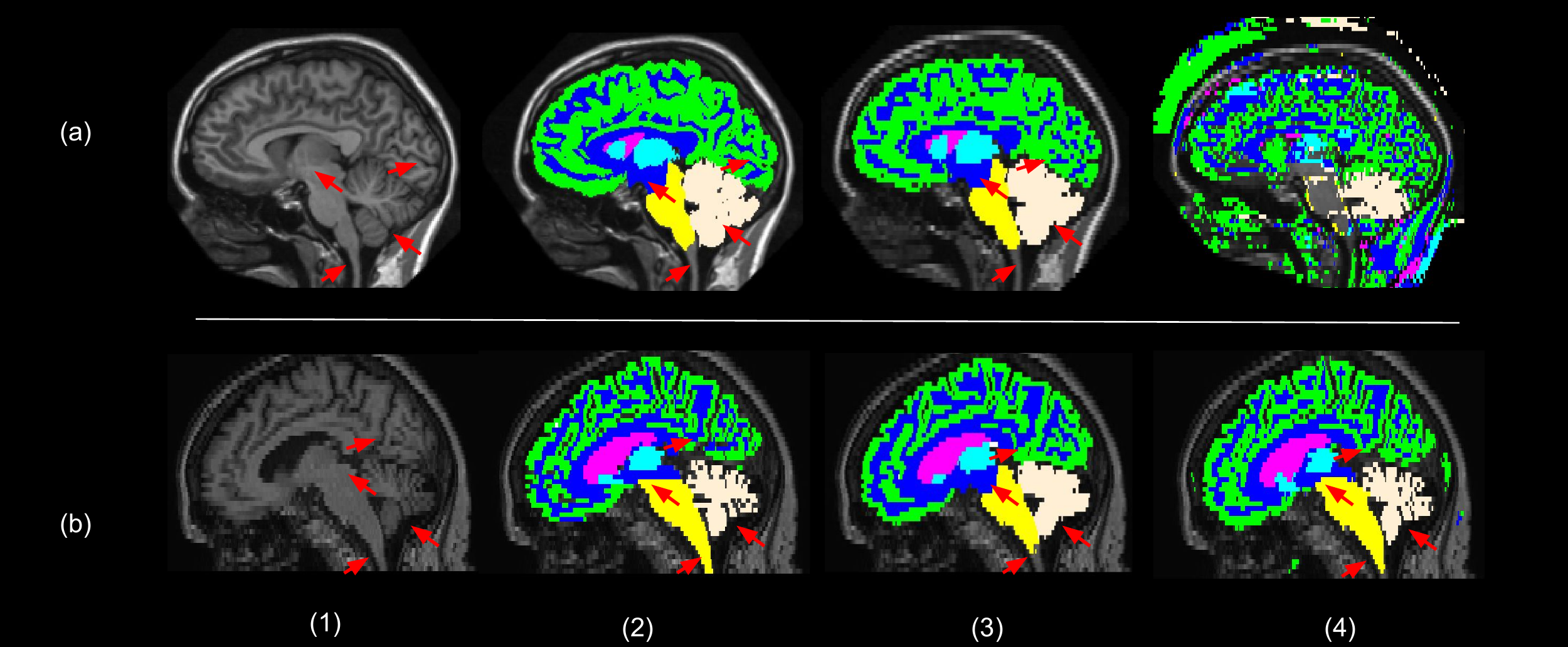}}
\end{figure}

\section{Conclusion}
We propose a joint model learned from hetero-modal datasets with disjoint heterogeneous annotations. Our approach is mathematically grounded, conceptually simple, new and relies on reasonable assumptions. We validated our approach by comparing our joint model with single-task learning models. We show that similar performance can be achieved for the tissue segmentation and lesion segmentation in comparison to task-specific baselines.
Moreover, our model achieves comparable performance to a model trained on a small fully-annotated joint dataset. 
Our work shows that the knowledge learnt from one modality is preserved when more modalities are used as input.
In the future, we will evaluate our approach on datasets with annotations protocols showing less variability.
Furthermore, exploitation of recent techniques for domain adaptation could help us bridge the gap and
improve the performance by helping to better satisfy some of our assumptions. Finally, we plan to integrate uncertainty measures in our framework as a future work.
As one of the first work to methodologically address the problem of joint learning from hetero-modal datasets, we believe that our approach will help DNN make further impact in clinical scenarios.

\midlacknowledgments{This work was supported by the Wellcome Trust [203148/Z/16/Z] and the Engineering and Physical
Sciences Research Council (EPSRC) [NS/A000049/1].  TV is supported by a Medtronic / Royal Academy of Engineering Research Chair [$RCSRF1819\backslash7\backslash34$].}

\bibliography{dorent19}

\begin{thebibliography}{22}
\providecommand{\natexlab}[1]{#1}
\providecommand{\url}[1]{\texttt{#1}}
\expandafter\ifx\csname urlstyle\endcsname\relax
  \providecommand{\doi}[1]{doi: #1}\else
  \providecommand{\doi}{doi: \begingroup \urlstyle{rm}\Url}\fi

\bibitem[Ashburner and Friston(2000)]{Ashburner:NeuroImage:2000}
John Ashburner and Karl~J. Friston.
\newblock Voxel-based morphometry -- the methods.
\newblock \emph{{Neuroimage}}, 11\penalty0 (6):\penalty0 805--821, 2000.

\bibitem[Barkhof and Scheltens(2002)]{Barkhof:CerebDis:2002}
Frederik Barkhof and Philip Scheltens.
\newblock Imaging of white matter lesions.
\newblock \emph{Cerebrovascular Diseases}, 13(Suppl 2):\penalty0 21--30, 2002.

\bibitem[Battaglini et~al.(2012)Battaglini, Jenkinson, and
  De~Stefano]{Battaglini:HBM:2012}
Marco Battaglini, Mark Jenkinson, and Nicola De~Stefano.
\newblock Evaluating and reducing the impact of white matter lesions on brain
  volume measurements.
\newblock \emph{Human Brain Mapping}, 33\penalty0 (9):\penalty0 2062--2071,
  2012.

\bibitem[Bottou et~al.(2018)Bottou, Curtis, and Nocedal]{Bottou:SIAMRev:2018}
L\'eon Bottou, Frank~E. Curtis, and Jorge Nocedal.
\newblock Optimization methods for large-scale machine learning.
\newblock \emph{{SIAM} Review}, 60\penalty0 (2):\penalty0 223--311, 2018.

\bibitem[Bragman et~al.(2018)Bragman, Tanno, Eaton-Rosen, Li, Hawkes, Ourselin,
  Alexander, McClelland, and Cardoso]{Bragman:MICCAI:2018}
Felix J.~S. Bragman, Ryutaro Tanno, Zach Eaton-Rosen, Wenqi Li, David~J.
  Hawkes, Sebastien Ourselin, Daniel~C. Alexander, Jamie~R. McClelland, and
  M.~Jorge Cardoso.
\newblock Uncertainty in multitask learning: Joint representations for
  probabilistic {MR}-only radiotherapy planning.
\newblock In \emph{Proceedings of the 21st International Conference on Medical
  Image Computing and Computer Assisted Intervention ({MICCAI}'18)}, pages
  3--11, 2018.

\bibitem[Cardoso et~al.(2015)Cardoso, Modat, Wolz, Melbourne, Cash, Rueckert,
  and Ourselin]{Cardoso:TMI:2015}
M.~Jorge Cardoso, Marc Modat, Robin Wolz, Andrew Melbourne, David Cash, Daniel
  Rueckert, and S\'ebastien Ourselin.
\newblock Geodesic information flows: Spatially-variant graphs and their
  application to segmentation and fusion.
\newblock \emph{{IEEE} Transactions on Medical Imaging}, 34\penalty0
  (9):\penalty0 1976--1988, September 2015.

\bibitem[Csurka(2017)]{Csurka:DACVARevChap:2017}
Gabriela Csurka.
\newblock \emph{A comprehensive survey on domain adaptation for visual
  applications}, pages 1--35.
\newblock Springer International Publishing, 2017.

\bibitem[Gibson et~al.(2018)Gibson, Li, Sudre, Fidon, Shakir, Wang,
  Eaton-Rosen, Gray, Doel, Hu, Whyntie, Nachev, Modat, Barratt, Ourselin,
  Cardoso, and Vercauteren]{Gibson:CMPB:2018}
Eli Gibson, Wenqi Li, Carole Sudre, Lucas Fidon, Dzhoshkun~I. Shakir, Guotai
  Wang, Zach Eaton-Rosen, Robert Gray, Tom Doel, Yipeng Hu, Tom Whyntie,
  Parashkev Nachev, Marc Modat, Dean~C. Barratt, S\'ebastien Ourselin, M.~Jorge
  Cardoso, and Tom Vercauteren.
\newblock {NiftyNet}: a deep-learning platform for medical imaging.
\newblock \emph{Computer Methods and Programs in Biomedicine}, 158:\penalty0
  113--122, 2018.

\bibitem[Havaei et~al.(2016)Havaei, Guizard, Chapados, and
  Bengio]{Havaei:MICCAI:2016}
Mohammad Havaei, Nicolas Guizard, Nicolas Chapados, and Yoshua Bengio.
\newblock {HeMIS}: Hetero-modal image segmentation.
\newblock In \emph{Proceedings of the 19th International Conference on Medical
  Image Computing and Computer Assisted Intervention ({MICCAI}'16)}, pages
  469--477, 2016.

\bibitem[He et~al.(2015)He, Zhang, Ren, and Sun]{He:ICCV:2015}
Kaiming He, Xiangyu Zhang, Shaoqing Ren, and Jian Sun.
\newblock Delving deep into rectifiers: Surpassing human-level performance on
  imagenet classification.
\newblock In \emph{Proceedings of the 25th International Conference on Computer
  Vision ({ICCV}'15)}, pages 1026--1034, December 2015.

\bibitem[Iglesias and Sabuncu(2015)]{Iglesias:MedIA:2015}
Juan~Eugenio Iglesias and Mert~R Sabuncu.
\newblock Multi-atlas segmentation of biomedical images: a survey.
\newblock \emph{Medical Image Analysis}, 24\penalty0 (1):\penalty0 205--219,
  August 2015.

\bibitem[Jenkinson et~al.(2012)Jenkinson, Beckmann, Behrens, Woolrich, and
  Smith]{Jenkinson:NeuroImage:2012}
Mark Jenkinson, Christian~F. Beckmann, Timothy~E.J. Behrens, Mark~W. Woolrich,
  and Stephen~M. Smith.
\newblock {FSL}.
\newblock \emph{{Neuroimage}}, 62\penalty0 (2):\penalty0 782--790, 2012.

\bibitem[Kamnitsas et~al.(2017)Kamnitsas, Baumgartner, Ledig, Newcombe,
  Simpson, Kane, Menon, Nori, Criminisi, Rueckert,
  et~al.]{Kamnitsas:MICCAI:2017}
Konstantinos Kamnitsas, Christian Baumgartner, Christian Ledig, Virginia
  Newcombe, Joanna Simpson, Andrew Kane, David Menon, Aditya Nori, Antonio
  Criminisi, Daniel Rueckert, et~al.
\newblock Unsupervised domain adaptation in brain lesion segmentation with
  adversarial networks.
\newblock In \emph{Proceedings of the 20th International Conference on Medical
  Image Computing and Computer Assisted Intervention ({MICCAI}'17)}, pages
  597--609, 2017.

\bibitem[Kendall and Gal(2017)]{Kendall:NIPS:2017}
Alex Kendall and Yarin Gal.
\newblock What uncertainties do we need in bayesian deep learning for computer
  vision?
\newblock In \emph{Proceedings of Advances in Neural Information Processing
  Systems 30 (NIPS 2017)}, pages 5574--5584, 2017.

\bibitem[Kim et~al.(2018)Kim, Choi, Oh, Yoon, and Kweon]{Kim:WACV:2018}
Dong-Jin Kim, Jinsoo Choi, Tae-Hyun Oh, Youngjin Yoon, and In~So Kweon.
\newblock Disjoint multi-task learning between heterogeneous human-centric
  tasks.
\newblock In \emph{Proceedings of the 2018 {IEEE} Winter Conference on
  Applications of Computer Vision ({WACV}) (2018)}, pages 1699--1708, 2018.

\bibitem[Kingma and Ba(2014)]{Kingma:arXiv:2014}
Diederik~P. Kingma and Jimmy Ba.
\newblock Adam: {A} method for stochastic optimization, 2014.
\newblock \href{http://arxiv.org/abs/1412.6980}{arXiv:1412.6980}.

\bibitem[Kosub(2018)]{Kosub:PRL:2018}
Sven Kosub.
\newblock A note on the triangle inequality for the {Jaccard} distance.
\newblock \emph{Pattern Recognition Letters}, December 2018.

\bibitem[Li et~al.(2017)Li, Wang, Fidon, Ourselin, Cardoso, and
  Vercauteren]{Li:IPMI:2017}
Wenqi Li, Guotai Wang, Lucas Fidon, Sebastien Ourselin, M.~Jorge Cardoso, and
  Tom Vercauteren.
\newblock On the compactness, efficiency, and representation of {3D}
  convolutional networks: Brain parcellation as a pretext task.
\newblock In \emph{Proceedings of Information Processing in Medical Imaging
  ({IPMI}'17)}, pages 348--360, 2017.

\bibitem[Li and Hoiem(2017)]{Li:PAMI:2017}
Zhizhong Li and Derek Hoiem.
\newblock Learning without forgetting.
\newblock \emph{{IEEE} Transactions on Pattern Analysis and Machine
  Intelligence}, 40\penalty0 (12):\penalty0 2935--2947, November 2017.

\bibitem[Milletari et~al.(2016)Milletari, Navab, and
  Ahmadi]{Milletari:3DV:2016}
Fausto Milletari, Nassir Navab, and Seyed-Ahmad Ahmadi.
\newblock {V-Net}: Fully convolutional neural networks for volumetric medical
  image segmentation.
\newblock In \emph{Proceedings of the 2016 Fourth International Conference on
  {3D} Vision ({3DV})}, pages 565--571, 2016.

\bibitem[Sdika and Pelletier(2009)]{Sdika:HBM:2009}
Micha\"el Sdika and Daniel Pelletier.
\newblock Nonrigid registration of multiple sclerosis brain images using lesion
  inpainting for morphometry or lesion mapping.
\newblock \emph{Human Brain Mapping}, 30\penalty0 (4):\penalty0 1060--1067,
  2009.

\bibitem[Ulyanov et~al.(2016)Ulyanov, Vedaldi, and
  Lempitsky]{Ulyanov:arXiv:2016}
Dmitry Ulyanov, Andrea Vedaldi, and Victor~S. Lempitsky.
\newblock Instance normalization: The missing ingredient for fast stylization,
  2016.
\newblock \href{http://arxiv.org/abs/1607.08022}{arXiv:1607.08022}.

\end{thebibliography}

\end{document}